\documentclass[12pt]{article}
\usepackage{amsmath}
\usepackage{graphicx}
\usepackage{color}
\usepackage{orcidlink}
\usepackage{hyperref}
\hypersetup{colorlinks=true, linkcolor=blue, citecolor=blue, urlcolor=blue}
\usepackage{subcaption}
\RequirePackage[numbers,sort&compress]{natbib}
\begin{document}
\baselineskip=18pt

\begin{center}
{\large{\bf Effects of gravity rainbow on scalar bosonic and oscillator fields in Bonnor-Melvin space-time with a cosmological constant }}
\end{center}

\vspace{0.3cm}

\begin{center}
    {\bf Faizuddin Ahmed\orcidlink{0000-0003-2196-9622}}\footnote{\bf faizuddinahmed15@gmail.com}\\
    \vspace{0.1cm}
    {\it Department of Physics, University of Science \& Technology Meghalaya, Ri-Bhoi, Meghalaya, 793101, India}\\
    \vspace{0.3cm}
    {\bf Abdelmalek Bouzenada\orcidlink{0000-0002-3363-980X}}\footnote{\bf abdelmalekbouzenada@gmail.com (Corresp. author)}\\
    \vspace{0.1cm}
    {\it  Laboratory of theoretical and applied Physics, Echahid Cheikh Larbi Tebessi University 12001, Algeria}\\
\end{center}

\vspace{0.3cm}

\begin{abstract}
In this paper, we explore the relativistic quantum motion of spin-zero scalar charge-free particles influenced by rainbow gravity's (RG's) in Bonnor-Melvin magnetic space-time, a four-dimensional solution featuring a positive cosmological constant. We solve the Klein-Gordon equation in this scenario by using two sets of rainbow function: (i) $f(\chi)=\frac{1}{(1-\beta\,\chi)}$,\,\, $h(\chi)=1$ and (ii) $f(\chi)=\frac{1}{(1-\beta\,\chi)}=h(\chi)$, where $0 < \chi\left(=\frac{|E|}{E_p}\right) \leq 1$ with $E_p$ being the Planck's energy, $E$ is the particle's energy, and $\beta$ is the rainbow parameter. Furthermore, we study the relativistic quantum oscillator through the Klein-Gordon oscillator equation in the same Bonnor-Melvin magnetic space-time under RG effects. Employing the first pair of rainbow function, we obtain an approximate eigenvalue solution of the quantum oscillator fields. Notably, we demonstrate that the relativistic energy profile of scalar and oscillator particles are influenced by the topology of the geometry and the cosmological constant which is related with the magnetic field strength lies along the symmetry axis. Additionally, we see the impact of rainbow parameter on these approximate relativistic energy profiles in both quantum systems.
\end{abstract}

\vspace{0.1cm}


\vspace{0.1cm}


\section{Introduction}\label{sec:1}

Embarking on a profound exploration of the intricate interplay between gravitational forces and the dynamics of quantum mechanical systems is an intellectually stimulating pursuit. Einstein's revolutionary general theory of relativity (GR) skillfully conceptualizes gravity as an intrinsic geometric feature of space-time \cite{k1}. This groundbreaking theory unravels a captivating connection between space-time curvature and the manifestation of classical gravitational fields, providing precise predictions for phenomena such as gravitational waves \cite{k2}, the observation of black hole shadows \cite{k3,t1,t2}, gravitational lensing, and gravitational redshift. Simultaneously, the framework of quantum mechanics (QM) \cite{k4} offers invaluable insights into the subtle behaviors of particles at the microscopic scale. The convergence of these two fundamental theories invites us to delve into the profound mysteries situated at the intersection of the macroscopic domain governed by gravity and the quantum intricacies of the subatomic realm.

Until a viable theory of quantum gravity (QG) is established, physicists are compelled to employ semi-classical approaches to quantum gravity. These approaches, while not providing a complete solution, prove beneficial in describing phenomena associated with extremely high energy physics and the early universe \cite{k5,k6,k7,k8,k9,k10,kk10,k11,k12}. An illustrative instance of such a phenomenological or semi-classical approach to quantum gravity involves the infringement of Lorentz invariance. This entails deviating from the ordinary relativistic dispersion relation, brought about by modifying the physical energy and momentum at the Planck scale \cite{k13}. This departure from the dispersion relation has found application in diverse realms, including space-time foam models \cite{k14}, loop quantum gravity \cite{k15}, spontaneous symmetry breaking of Lorentz invariance in string field theory \cite{k16}, spin networks \cite{k17}, discrete space-time \cite{k18}, and non-commutative geometry and Lorentz invariance violation \cite{k19}.

In Refs. \cite{k8, pp2}, the authors have introduced a modified framework known as double special relativity (DSR) theory, which is based on a deformed energy-momentum dispersion relation. In this theory, besides the speed of light being the maximum attainable velocity, there exists another fundamental constant, the Planck energy $E_p$, which sets the maximum energy scale in nature. This presents a different perspective on special relativity within the realm of microscopic physics. This theory has been further extended to include curved space-time, resulting in what is known as gravity's rainbow (RG) \cite{k8,k9,kk10}. In gravity's rainbow, the geometry of space-time becomes energy-dependent, meaning observers with different energies would perceive different geometries of space-time. As a consequence, a family of energy-dependent metrics, known as rainbow metrics, is used to describe the space-time geometry, which differs from the predictions of general relativity.

The relativistic energy-momentum relation in special theory of relativity (STR) is given by (choosing $c=1$):
\begin{equation}
    E^2-p^2=M^2, \label{ab1}
\end{equation}
where $\vec{p}$ is the relativistic momentum vector, $M$ is the rest mass, and $E$ is the relativistic particle's energy. Based on the nonlinearity of Lorentz transformations, the energy-momentum deformation relation can be rewritten as
\begin{equation}
    E^2\,f^2\left(E/E_p\right)-p^2\,h^2\left(E/E_p\right)=M^2, \label{ab2}
\end{equation}
where $E_p$ is the Planck's energy, and $f\left(E/E_p\right)$ and $h\left(E/E_p\right)$ are the correction terms known as rainbow functions. These functions are required to satisfy the following relationship
\begin{equation}
    \lim_{E/E_{p}\to 0} f\left(E/E_p\right) \to 1,\quad\quad \lim_{E/E_{p}\to 0}\,h\left(E/E_p\right) \to 1.\label{ab3} 
\end{equation}
Therefore, one can see that the deformed energy-momentum dispersion relation, Eq. (\ref{ab2}), will converge to the classical one, Eq. (\ref{ab1}), when the energy of the test particle is much lower than the Planck energy $E_p$.

Scientists have focused on applications of RG's across various realms of physics. These explorations span a diverse array of topics, encompassing the isotropic quantum cosmological perfect fluid model within the framework of RG \cite{k20}, adaptation of Friedmann-Robertson-Walker universe in the context of Einstein-massive RG \cite{k21}, thermodynamics governing black holes \cite{k22}, geodesic structure characterizing the Schwarzschild black hole \cite{k23,k24}, and examination of massive scalar field in the presence of Casimir effect \cite{k25}. Equally intriguing are inquiries into the impact of RG on equilibrium configurations elucidated by Tolman-Oppenheimer-Volkoff equation \cite{k26}, interplay between RG and Hořava-Lifshitz gravity \cite{k27}, dynamic interplay of topology changes and emergence of electric or magnetic charges due to quantum fluctuations in the context of RG \cite{k28,k29}, effects arising from the combination of $f(R)$ gravity theory with RG on the computation of the induced cosmological constant \cite{k30}, black hole entropy utilizing the brick wall model in conjunction with a rainbow metric \cite{k31}, zero-point energy and ultraviolet divergences through judicious selection of rainbow functions \cite{k32,k33,k34} etc.. In addition, this exploration extends further to encompass topics, such as quantum cosmology \cite{k35}, intricate geometries of wormholes in both cis-Planckian and trans-Planckian regimes employing rainbow functions \cite{k36}, identification of temporal divergences for ingoing observers in RG beyond the Planck scale \cite{k37}, dynamics of gravitational collapse within the framework of RG \cite{k38}, and some other investigations under RG's were reported in \cite{BEP1,BEP2,BEP3,BEP4,BEP5,BEP6,BEP7}. These scholarly articles collectively underscore the growing interest in the semi-classical approach to RG's and highlight the potential applications of the findings presented there. In this paper, the Klein-Gordon (KG) equation and its oscillator version is investigated within a magnetic space-time incorporating RG and topological defects.

In recent years, a great interest has emerged in the exploration of magnetic fields, spurred by discoveries of systems with remarkably strong fields like magnetars \cite{kk13, kk14} and occurrences in heavy ion collisions \cite{kk15, kk16, kk17}. The integration of the magnetic field within the framework of general relativity prompts intriguing questions, considering various existing solutions to the Einstein-Maxwell equations, including the Manko solution \cite{kk18, kk19}, the Bonnor-Melvin universe \cite{kk20, kk21, MZ2, MZ3}, and a recent space-time model in \cite{kk22} that introduces the cosmological constant into the Bonnor-Melvin framework. Shifting focus to the intersection of general relativity and quantum physics raises another critical consideration: the potential interrelation of these two theories and its relevance. Numerous studies have tackled this inquiry, often relying on the Klein-Gordon and Dirac equations within curved space-times \cite{kk23, kk24}. This exploration spans diverse scenarios, from particles in Schwarzschild \cite{kk25}, Kerr black holes \cite{kk26}, cosmic string backgrounds \cite{kk27,kk28,FA}, quantum oscillators \cite{kk32, kk35, kk35-1, kk35-2, kk35-3, AG}, the Casimir effect \cite{kk36, kk37}, and particles within the Hartle-Thorne space-time \cite{kk38}. These investigations, alongside others \cite{kk39, kk40} have provided valuable insights into how quantum systems respond to the arbitrary geometries of space-time. Consequently, an intriguing avenue of study involves examining quantum particles within a space-time influenced by a magnetic field. For instance, in \cite{kk42}, the quantum dynamics of Dirac particles were explored in the Melvin metric.

\begin{figure}[ht!]
    \centering
    \includegraphics[width=0.6\linewidth]{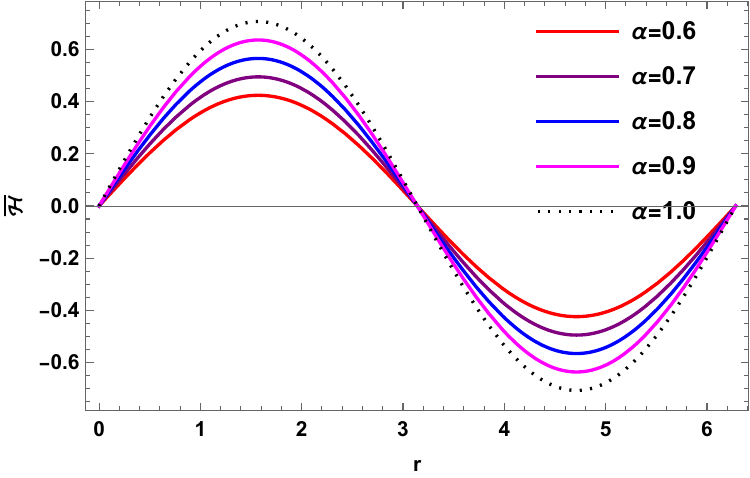}
    \caption{The behavior of the magnetic field strength as the cosmic string parameter $\alpha$ varies. The dotted line represents the case where $\alpha=1$.}
    \label{figure:0}
\end{figure}

{\color{black} In Ref. \cite{kk22}, Zofka presented a solution of field equations, a generalization of the well-known Bonnor-Melvin solution of the Einstein-Maxwell equations. This solution is a cylindrically symmetric space-time featuring a magnetic field aligned with the axis and a
 nonvanishing cosmological constant. This space-time we called Bonnor-Melvin universe with a positive cosmological constant. The line-element describing this space-time is given by (see Eq. (30) in Ref. \cite{kk22})
\begin{equation}
ds^{2}=-dt^2+dr^2+dz^2+\sigma^2\,\sin^2 (\sqrt{2\,\Lambda}\,r)\,d\phi^2,\label{a0}
\end{equation} 
where $\sigma>0$ is an integration constant and $\Lambda>0$ denotes the cosmological constant. The magnetic field strength is given by (see Eq. (31) in Ref. \cite{kk22})
\begin{equation}
    \mathcal{H} (r)=\sqrt{\Lambda}\,\sigma\,\sin (\sqrt{2\,\Lambda}\,r).\label{a00}
\end{equation}

By rescaling both $r \to \frac{r'}{\sqrt{2\,\Lambda}}$\quad and \quad $\phi \to \frac{\phi'}{\sqrt{2\,\Lambda}}$ (instead of $\phi \to \frac{\phi'}{\sigma\,\sqrt{2\,\Lambda}}$ as done in Ref. \cite{kk22}), and changing $\sigma \to \alpha$,} the above line-element (\ref{a00}) can be rewritten as (dropping prime)
\begin{equation}
ds^{2}=-dt^2+dz^2+\frac{1}{2\,\Lambda}\,\Big(dr^2+\alpha^2\,\sin^2 r\,d\phi^2\Big),\label{a1}
\end{equation}
where $\alpha$ represents a cosmic string parameter since it produces an angular deficit by an amount $\Delta\phi=2\,\pi\,(1-\alpha)$. {\color{black} The magnetic field strength for the space-time (\ref{a1}) is given by: 
\begin{equation}
    \bar{\mathcal{H}}(r)=\frac{\alpha\,\sin r}{\sqrt{2}}=\alpha\,\mathcal{H}(r)\label{aa1}
\end{equation} 
which is scaled by a factor of $\alpha$ compared to the field strength ($\mathcal{H}$) derived in \cite{kk22}. This result indicates that the magnetic field strength is influenced by the cosmic string parameter $\alpha$. The behaviour of the field strength is shown in Figure \ref{figure:0}.}

Recently, the Bonnor-Melvin solution with a cosmological constant, given by Eq. (\ref{a1}), has garnered significant attention in quantum mechanical systems (see, for example, \cite{AB2,AB3,AB4}). In the presence of rainbow gravity effects, this magnetic solution is described by the following line element, where the modifications $dt \to \frac{dt}{f(\chi)}$ and $dx^{i} \to \frac{dx^i}{h(\chi)}$ are applied:
\begin{equation}
ds^{2}=-\frac{dt^2}{f^2(\chi)}+\frac{1}{h^2(\chi)}\,\Big[dz^2+\frac{1}{2\,\Lambda}\,\Big(dr^2+\alpha^2\,\sin^2 r\,d\phi^2\Big)\Big],\label{a2}
\end{equation}
where $f(\chi)$ and $h=h(\chi)$ are the rainbow functions and the dimensionless parameter $\chi=\frac{|E|}{E_p}$, with $E_p$ being the Planck's energy and $E$ is the particle's energy. This dimensionless parameter lies in the interval $0 \leq \chi < 1$. In the limit $f(\chi) \to 1$ and $h(\chi) \to 1$, one will get back the cosmological Bonnor-Melvin universe with a cosmic string given in (\ref{a1}).

The covariant metric tensor $g_{\mu\nu}$ and its contravariant form $g^{\mu\nu}$ for the space-time (\ref{a2}) are given by
\begin{eqnarray}
g_{\mu\nu}=\begin{pmatrix}
-\frac{1}{f^2} & 0 & 0 & 0\\
0 & \frac{1}{2\,\Lambda\,h^2} & 0 & 0\\
0 & 0 & \frac{\alpha^2\,\sin^2 r}{2\,\Lambda\,h^2} & 0\\
0 & 0 & 0 & \frac{1}{h^2}
\end{pmatrix},\quad 
g^{\mu\nu}=\begin{pmatrix}
-f^2 & 0 & 0 & 0\\
0 & 2\,\Lambda\,h^2 & 0 & 0\\
0 & 0 & \frac{2\,\Lambda\,h^2}{\alpha^2\,\sin^2 r} & 0\\
0 & 0 & 0 & h^2
\end{pmatrix}. \label{a3}
\end{eqnarray}
The determinant of the metric tensor for the space-time (\ref{a2}) is given by 
\begin{equation}
    det\,(g_{\mu\nu})=g=-\frac{\alpha^2\,\sin^2 r}{4\,\Lambda^2\,h^6\,f^2}. \label{a4}
\end{equation}

Our motivation is to study the relativistic quantum motions of scalar particles under the influence of RG's in the background of a Bonnor-Melvin solution. This is a four-dimensional space-time with a cosmological constant attached with the magnetic field strength along $z$-direction. We derive the radial equation in this magnetic universe background and obtain approximate as well as exact analytical eigenvalue solutions by choosing two pairs of rainbow functions: ((i) $f(\chi)=\frac{1}{1-\beta\,\chi}$,\quad $h(\chi)=1$ with $\chi=E/E_p$ \cite{ZWF} and (ii) $f(\chi)=\frac{1}{1-\beta\,\chi}=h(\chi)$ \cite{k8,k9,kk10,SHH}. It is worth noting that the chosen pair of rainbow functions is a suitable set used to address the horizon problem. These rainbow functions were discussed in Refs. \cite{k8,k9,kk10,SHH} in the context of studying possible nonsingular universe solutions. By ensuring a constant light velocity, they may offer a resolution to the horizon problem. Our objective is to utilize this pair of rainbow functions within the framework of quantum field theory in curved space-time. Afterwards, we study the relativistic quantum oscillator fields in the background of same geometry with RG function $f(\chi)=\frac{1}{1-\beta\,\chi}$,\quad $h(\chi)=1$ and obtain the approximate energy profiles. In fact, we show that the energy profiles of scalar and oscillator fields are influenced by the topology of the space-time and the cosmological constant. Furthermore. rainbow parameter modifies the energy spectrum and shifted the results.     

This paper is summarized as follows: In section \ref{sec:2}, we study the Klein-Gordon equation in the background of Bonnor-Melvin magnetic universe with a cosmological constant in the presence of RG's. We derive the radial equation obtain the approximate as well as analytical solutions by choosing two pairs of rainbow functions. In section \ref{sec:3}, we study the relativistic quantum oscillator fields in the same magnetic universe background and obtain the approximate eigenvalue solution by choosing on pair of rainbow functions. In section \ref{sec:4}, we present our results and discussion. Throughout this paper, {\color{black}{we use the system of units, where $\hbar=c=G=1$}}.

\section{Effects of RG on scalar bosons in BM-universe with a cosmological constant }\label{sec:2}

In this part, we study the relativistic quantum motions of charge-free scalar particles under the influence of RG's within the context of a cosmological space-time. In our interest, we consider the Bonnor-Melvin magnetic space-time and solve the Klein-Gordon equation in the presence of RG's. Therefore, we begin this section by writing the relativistic wave equation describing the quantum motions of charge-free scalar particles is described by the following equation \cite{kk32, kk35, kk35-1, kk35-2, kk35-3, WG}
\begin{eqnarray}
    \Bigg[-\frac{1}{\sqrt{-g}}\,\partial_{\mu}\,\Big(\sqrt{-g}\,g^{\mu\nu}\,\partial_{\nu}\Big)+M^2\Bigg]\,\Psi=0,\label{b1}
\end{eqnarray}
where $M$ is the rest mass of the particles, $g$ is the determinant of the metric tensor $g_{\mu\nu}$ with its inverse $g^{\mu\nu}$. 

Therefore, expression the KG-equation (\ref{b1}) in the space-time background (\ref{a2}) and using (\ref{a3})--(\ref{a4}), we obtain the following differential equation
\begin{eqnarray}
    \Bigg[-f^2\,\frac{d^2}{dt^2}+2\,\Lambda\,h^2\,\Bigg\{\frac{d^2}{dr^2}+\frac{1}{\tan r}\,\frac{d}{dr}+\frac{1}{\alpha^2\,\sin^2 r}\,\frac{d^2}{d\phi^2}\Bigg\}+h^2\,\frac{d^2}{dz^2}-M^2\Bigg]\,\Psi (t, r, \phi, z)=0\,.\label{b2}
\end{eqnarray}

From the differential equation (\ref{b2}), one can see that the wave function $\Psi (t, r, \phi, z)$ is independent of the coordinates $t, \phi$ and $z$. Therefore, we choose the following ansatz for the wave function $\Psi (t, r, \phi, z)$ in terms of the radial function $\psi (r)$ as follows:
\begin{equation}
    \Psi (t, r, \phi, z) =\exp(-i\,E\,t)\,\exp(i\,m\,\phi)\,\exp(i\,k\,z)\,\psi(r), \label{b3}
\end{equation}
where $E$ is the particle's energy, $m=0,\pm\,1,\pm\,2,....$ are the eigenvalues of the angular quantum operator, and $k$ the wave-number which is arbitrary constant.

Substituting the total wave function (\ref{b3}) into the differential equation (\ref{b2}) and after separating the variables, we obtain the differential equations for $\psi(x)$ as follows:
\begin{equation}
    \psi''+\frac{1}{\tan r}\,\psi'+\Bigg[\frac{(f^2\,E^2-M^2)}{2\,\Lambda\,h^2}-\frac{k^2}{2\,\Lambda}-\frac{\iota^2}{\sin^2 r}\Bigg]\,\psi=0,\label{b4}
\end{equation}
where $\iota=\frac{|m|}{\alpha}$.

Now, we solve the above differential equation analytically and obtain the energy eigenvalue and wave function of the scalar particles. In order to solve Eq. (\ref{b4}), {\color{black} we consider an approximation such that $\sin r =\frac{1}{2\,i}\,(e^{i\,r}-e^{-i\,r})\approx r$, $\tan r \approx r$ which is valid for small value of $r<<1$. However, the exact solutions to differential equation (\ref{b4}) can be found through special functions, but we leave this for the broader readership to explore.} 

Therefore, by employing this approximation into the radial equation (\ref{b4}) results
\begin{equation}
    \psi''+\frac{1}{r}\,\psi'+\Bigg[\frac{(f^2\,E^2-M^2)}{2\,\Lambda\,h^2}-\frac{k^2}{2\,\Lambda}-\frac{\iota^2}{r^2}\Bigg]\,\psi=0.\label{c1}
\end{equation}
The above second-order differential equation is the Bessel differential equation form \cite{MA, GBA} whose solutions are well-known in the literature. We want a solution which is finite at the origin $r=0$ and this is possible only for the Bessel functions of the first kind given by 
\begin{equation}
    \psi (r)=c_1\,J_{\iota} (r\,\Theta), \label{c2}
\end{equation}
where $c_1$ is an arbitrary constant and $\Theta=\sqrt{\frac{(f^2\,E^2-M^2)}{2\,\Lambda\,h^2}-\frac{k^2}{2\,\Lambda}}$.

Asymptotic form of the Bessel function \cite{MA,GBA} is given by
\begin{equation}
    J_{\iota} (r\,\Theta) \propto \cos \Big(r\,\Theta-\frac{\iota\,\pi}{2}-\frac{\pi}{4}\Big)\label{c3}
\end{equation}
Now, we impose a hard wall confining potential condition which states that at some axial distance $r=r_0>0$, the radial wave function $\psi (r=r_0)=0$. This condition is also known as Dirichlet's boundary condition. Therefore, substituting (\ref{c3}) into the equation (\ref{c2}) and using this condition $\psi (r=r_0)=0$, we obtain the following relation of the energy spectrum given by
\begin{equation}
    f^2\,E^2=M^2+h^2\,\Bigg[k^2+\Lambda\,\Bigg(2\,n+\frac{3}{2}+\frac{|m|}{\alpha}\Bigg)^2\,\frac{\pi^2}{2\,r^2_{0}}\Bigg],\label{c4}
\end{equation}
where $n=0, 1,2,3,....$.

Now, we use two different types of rainbow function and obtain the compact expression of the energy eigenvalue.

\subsection{Rainbow functions $f(\chi)=\frac{1}{1-\beta\,\chi}$,\quad $h(\chi)=1$ with $\chi=\frac{E}{E_p}$.}

{\color{black} By substituting a pair of rainbow function given by $f(\chi)=\frac{1}{1-\beta\,\chi}$,\quad $h(\chi)=1$ into the expression (\ref{c4}), we finds the following eigenvalue equation}
\begin{equation}
    \frac{E^2}{\Big(1-\beta\,\frac{E}{E_p}\Big)^2}=M^2+\Bigg[k^2+\Lambda\,\Bigg(2\,n+\frac{3}{2}+\frac{|m|}{\alpha}\Bigg)^2\,\frac{\pi^2}{2\,r^2_{0}}\Bigg].\label{c44}
\end{equation}

Simplification of the above equation gives us the approximate energy eigenvalue associated with the mode $\{n, m\}$ given by 
\begin{equation}
    E_{n\,m}=\Bigg[\frac{\beta}{E_p}\pm \Bigg\{M^2+k^2+2\,\Lambda\,\Big(n+\frac{3}{4}+\frac{|m|}{2\,\alpha}\Big)^2\,\frac{\pi^2}{r^2_{0}}\Bigg\}^{-1/2}\Bigg]^{-1}.\label{c5}
\end{equation}

\begin{center}
\begin{figure}[ht!]
\begin{centering}
\subfloat[$\alpha=0.9$, $n=1=m$]{\centering{}\includegraphics[scale=0.47]{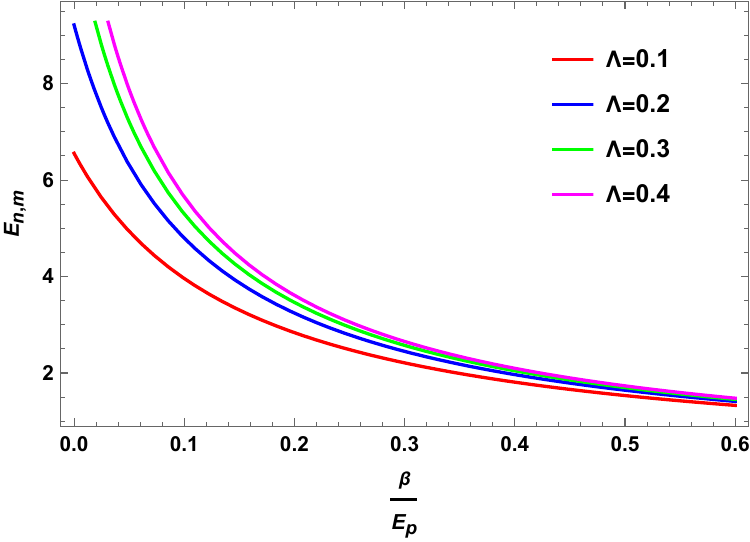}}\quad\quad\quad
\subfloat[$\Lambda=0.1$,$n=1=m$]{\centering{}\includegraphics[scale=0.47]{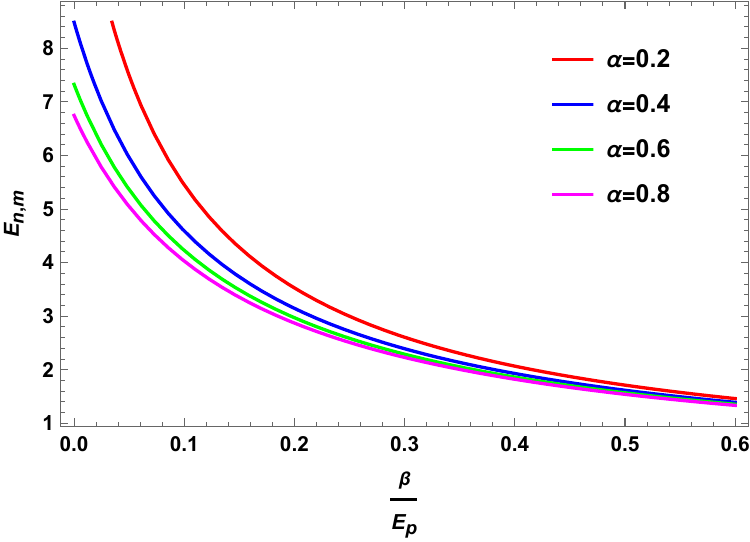}}\\
\subfloat[$\Lambda=0.1$, $\alpha=0.9$, $m=1$]{\centering{}\includegraphics[scale=0.47]{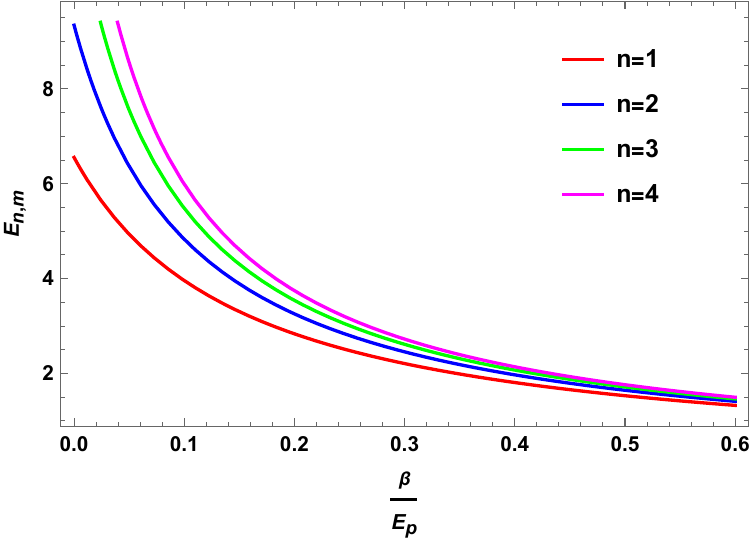}}\quad\quad\quad
\subfloat[$\alpha=0.9$, $m=1$]{\centering{}\includegraphics[scale=0.47]{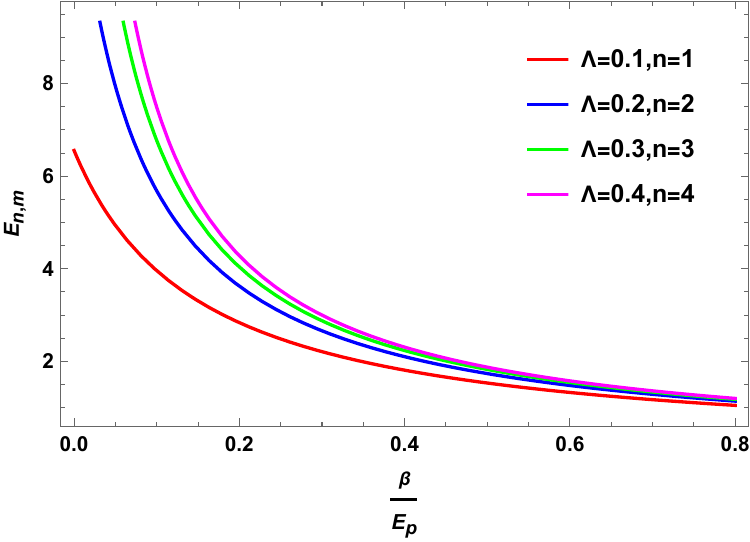}}
\centering{}\caption{The energy level $E^{+}_{n\,m}$ of Eq. (\ref{c5}) as a function of $\beta/E_p$ for different values of $(n,\Lambda,\alpha)$. Here, $M=1, k=0, r_0=0.5$.}
\label{figure:1}
\hfill\\
\subfloat[$\alpha=0.9$, $n=1=m$]{\centering{}\includegraphics[scale=0.47]{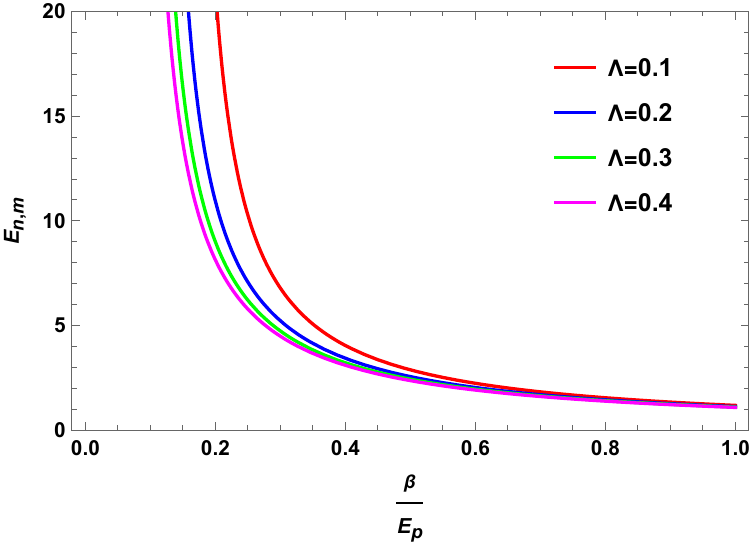}}\quad\quad\quad
\subfloat[$\Lambda=0.1$,$n=1=m$]{\centering{}\includegraphics[scale=0.47]{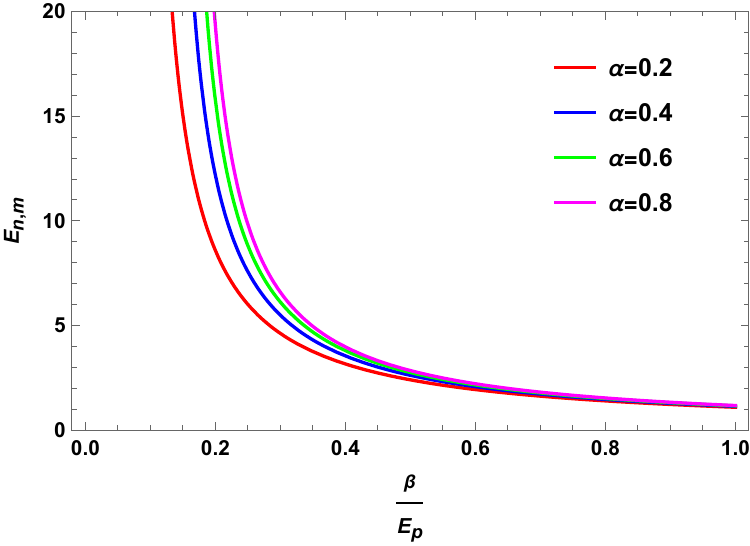}}\\
\subfloat[$\Lambda=0.1$, $\alpha=0.9$, $m=1$]{\centering{}\includegraphics[scale=0.47]{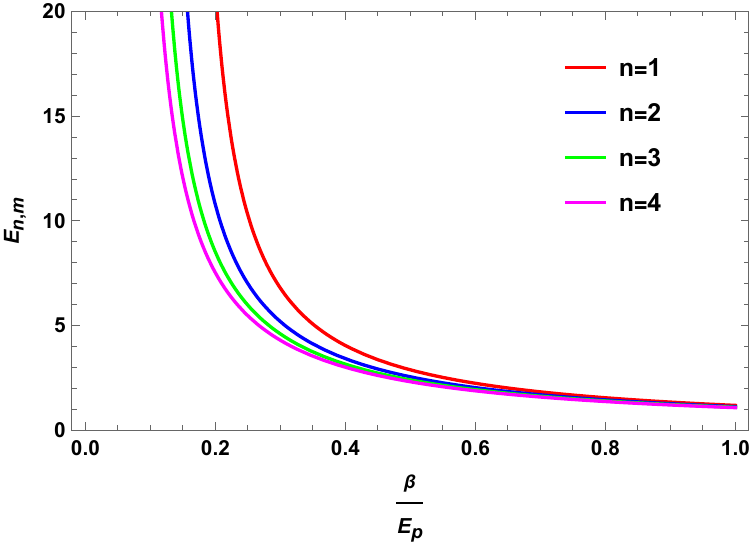}}\quad\quad\quad
\subfloat[$\alpha=0.9$, $m=1$]{\centering{}\includegraphics[scale=0.47]{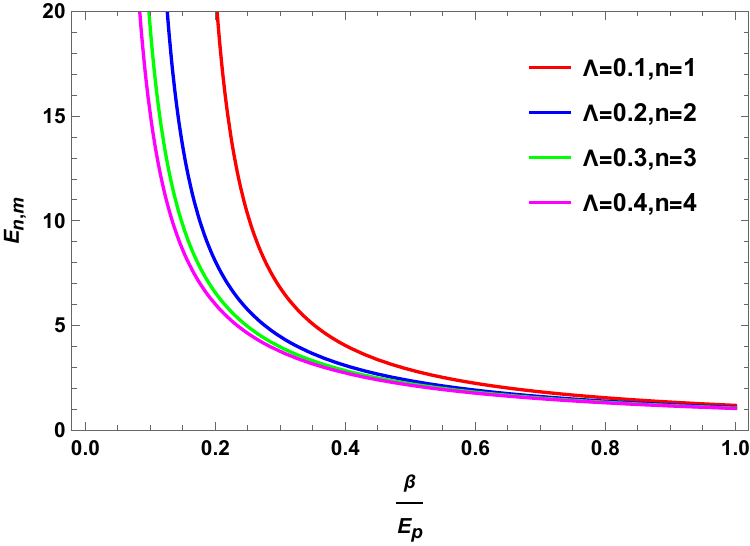}}
\centering{}\caption{The energy level $E^{-}_{n\,m}$ of Eq. (\ref{c5}) as a function of $\beta/E_p$ for different values of $(n,\Lambda,\alpha)$. Here, $M=1, k=0, r_0=0.5$.}
\label{figure:2}
\end{centering}
\end{figure}
\par\end{center}

{\color{black} Equation (\ref{c5}) is the approximate relativistic energy profile of a scalar particle in BM-universe with a cosmological constant under the influence of rainbow gravity.} We see that the relativistic energy spectrum is influenced by the topology of the geometry characterized by the parameter $\alpha$ and the cosmological constant $\Lambda$. Furthermore, the rainbow parameter $ \beta <1$ also modified the energy profiles and shifted the results.

We have presented Figures \ref{figure:1}--\ref{figure:2}, illustrating the energy spectrum $E_{n, m}$ plotted against $\beta/E_p$. These figures encompasses different values of the cosmological constant (Fig: 1(a) \& 2(a)), the topology parameter $\alpha$ (Fig: 1(b) \& 2(b)), the radial quantum number $n$ (Fig: 1(c) \& 2(c)), and both the cosmological constant $\Lambda$ and quantum number $n$ (Fig: 1(d) \& 2(d)). Notably, the nature of the spectrum is parabolic and exhibits an upward shift with increasing values of $\Lambda$ and $m$, while depicting a decreasing trend for $\alpha$. {\color{black} It is worth mentioning that in Figures \ref{figure:1}--\ref{figure:2}, as well as in other figures throughout the manuscript, we have set the values of various parameters in geometric units, where $c=1=\hbar$. Additionally, for $\beta/E_p = 0$, which implies $\beta=0$, the energy eigenvalue $E_{n\,m}$ remains finite and non-zero. Therefore, all figures in the manuscript depict the energy eigenvalue $E_{n\,m}$ as a function of $\beta/E_p$.}

{\color{black}
In the limit $\beta=0$, the rainbow functions become $f(\chi) \to 1$ and  $h(\chi) \to 1$ and hence, the geometry (\ref{a2}) reduces to the original one (\ref{a1}). In that case, the approximate relativistic energy eigenvalue from Eq. (\ref{c5}) will be
\begin{equation}
    E_{n\,m}=\pm \sqrt{M^2+k^2+2\,\Lambda\,\Big(n+\frac{3}{4}+\frac{|m|}{2\,\alpha}\Big)^2\,\frac{\pi^2}{r^2_{0}}}.\label{special}
\end{equation}

Equation (\ref{special}) is the approximate relativistic energy levels of spin-0 charge-free scalar particles within the context of Bonnor-Melvin-universe featuring a cosmological constant without rainbow gravity effects. 

Thus, by comparing Eqs. (\ref{c5}) and (\ref{special}), it becomes evident that the presence of rainbow gravity alters the relativistic energy levels of scalar quantum particles when compared to the scenario without this effect.}

\subsection{Rainbow functions $f(\chi)=\frac{1}{1-\beta\,\chi}=h(\chi)$}

By substituting a pair of rainbow function given by $f(\chi)=\frac{1}{1-\beta\,\chi}=h(\chi)$ into the expression (\ref{c4}) results
\begin{equation}
    \frac{E^2}{\Big(1-\beta\,\frac{E}{E_p}\Big)^2}=M^2+\frac{1}{\Big(1-\beta\,\frac{E}{E_p}\Big)^2}\,\Bigg[k^2+\Lambda\,\Bigg(2\,n+\frac{3}{2}+\frac{|m|}{\alpha}\Bigg)^2\,\frac{\pi^2}{2\,r^2_{0}}\Bigg].\label{c444}
\end{equation}
Simplification of the above equation gives us the approximate energy eigenvalue of scalar particles associated with the mode $\{n, m\}$ given by 
\begin{equation}
    E_{n\,m}=\frac{1}{\Big(1-\frac{M^2\,\beta^2}{E^2_{p}}\Big)}\,\Bigg[-\frac{\beta\,M^2}{E_p} \pm \sqrt{M^2+\Big(1-\frac{M^2\,\beta^2}{E^2_{p}}\Big)\,\Bigg\{k^2+2\Lambda\Bigg(n+\frac{3}{4}+\frac{|m|}{2\,\alpha}\Bigg)^2\,\frac{\pi^2}{r^2_{0}}\Bigg\}}\Bigg].\label{c6}
\end{equation}

Equation (\ref{c6}) is the approximate relativistic energy profile of scalar particles in BM-universe with a cosmological under the influence of rainbow gravity. We see that the relativistic energy spectrum is influenced by the topology of the geometry characterized by the parameter $\alpha$, and the cosmological constant $\Lambda$. {\color{black} In the limit $\beta = 0$, the approximate relativistic energy eigenvalue (\ref{c6}) reduces to the energy spectrum obtained in Eq. (\ref{special}).}

We present Figure \ref{figure:3}, illustrating the energy spectrum $E_{n, m} (+)$ (considering the positive sign within the brackets in the energy expression (\ref{c6})) plotted against $E_p/\beta$. The figure displays various values of the cosmological constant (Fig: 3(a)), the topology parameter $\alpha$ (Fig: 3(b)), and the radial quantum number $n$ (Fig: 3(c)), and both the cosmological constant $\Lambda$ and quantum number $n$ (Fig: 3(d)).

\begin{center}
\begin{figure}[ht!]
\begin{centering}
\subfloat[$\alpha=0.9$, $n=1=m$]{\centering{}\includegraphics[scale=0.55]{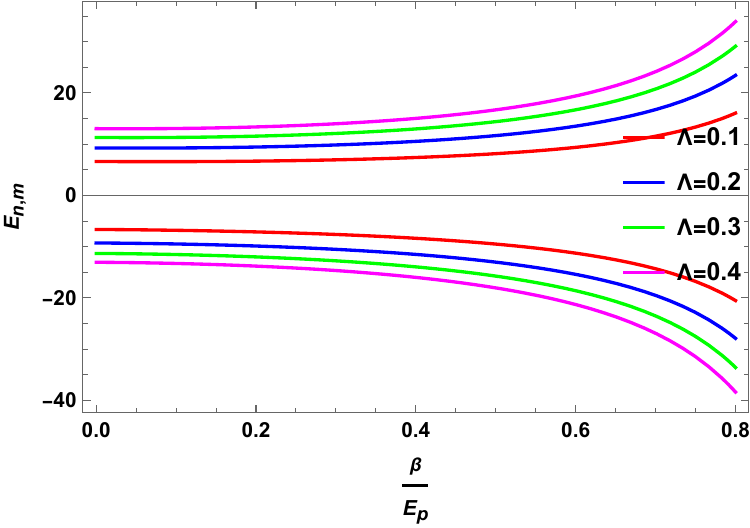}}\quad\quad\quad
\subfloat[$\Lambda=0.1$,$n=1=m$]{\centering{}\includegraphics[scale=0.55]{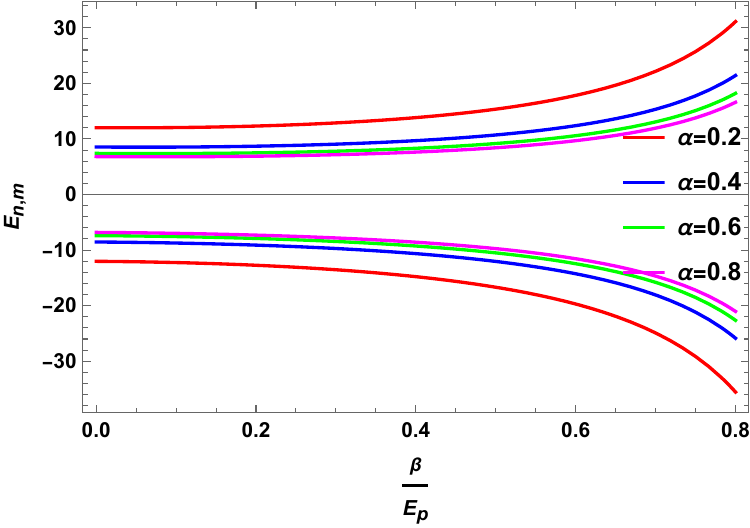}}\\
\subfloat[$\Lambda=0.1$, $\alpha=0.9$, $m=1$]{\centering{}\includegraphics[scale=0.55]{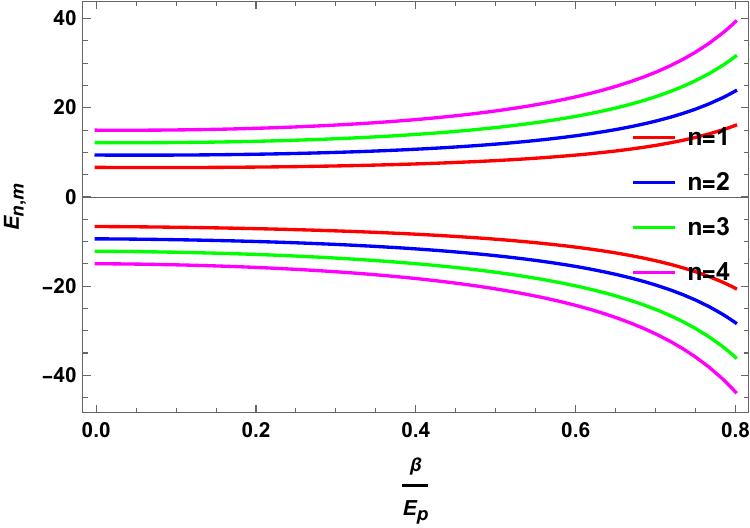}}\quad\quad\quad
\subfloat[$\alpha=0.9$, $m=1$]{\centering{}\includegraphics[scale=0.55]{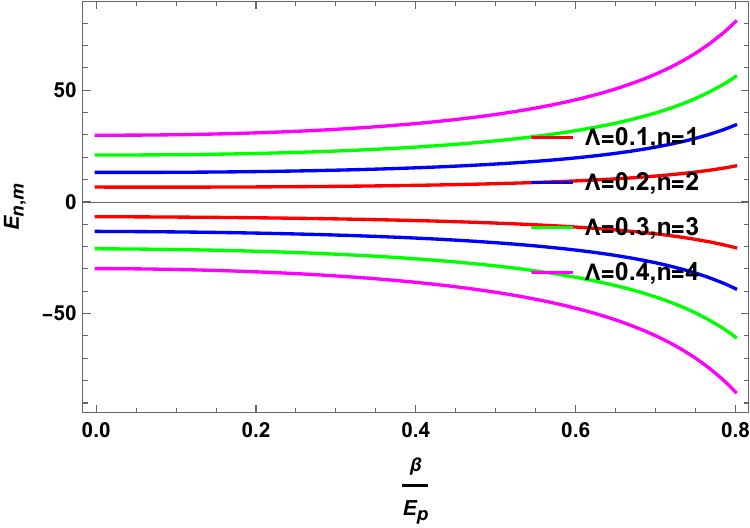}}
\centering{}\caption{ The energy spectrum $E_{n\,m}$ of equation (\ref{c6}), where parameters are set as $M=1, k=0, r_0=0.5$.}
\label{figure:3}
\end{centering}
\end{figure}
\par\end{center}

\section{Effects of RG on relativistic quantum oscillator field in BM-universe with a cosmological constant}\label{sec:3}

In this part, we study the relativistic quantum oscillator fields via the Klein-Gordon oscillator in the background of Bonnor-Melvin magnetic universe. This oscillator field is studied by replacing the momentum operator into the Klein-Gordon equation via $\partial_{\mu} \to (\partial_{\mu}+M\,\omega\,X_{\mu})$, where the four-vector $X_{\mu}=(0, r, 0, 0)$ and $\omega$ is the oscillator frequency. The relativistic quantum oscillator fields in curved space-times background have been investigated by numerous authors (see, \cite{kk32, kk35, kk35-1, kk35-2, kk35-3}).

Therefore, the relativistic wave equation describing the quantum oscillator fields is given by
\begin{eqnarray}
    \Bigg[\frac{1}{\sqrt{-g}}\,(\partial_{\mu}+M\,\omega\,X_{\mu})\,(\sqrt{-g}\,g^{\mu\nu})\,(\partial_{\nu}-M\,\omega\,X_{\nu})\Bigg]\,\Psi=M^2\,\Psi.  \label{e1}
\end{eqnarray}
Expressing the wave equation (\ref{e1}) in the magnetic universe background (\ref{a2}), we obtain
\begin{eqnarray}
    &&\Bigg[-f^2\frac{d^2}{dt^2}+2\Lambda h^2\Bigg\{\frac{d^2}{dr^2}+\frac{1}{\tan r}\frac{d}{dr}-M \omega-\frac{M \omega r}{\tan r}-M^2 \omega^2 r^2+\frac{1}{\alpha^2 \sin^2 r}\frac{d^2}{d\phi^2}\Bigg\}\nonumber\\
    &&+h^2 \frac{d^2}{dz^2}-M^2\Bigg]\Psi=0.\label{e2}
\end{eqnarray}
Substituting the wave function ansatz (\ref{b3}) into the above differential equation (\ref{e2}) results the following second-order differential equation form:
\begin{equation}
    \psi''+\frac{1}{\tan r}\,\psi'+\Bigg[\frac{(f^2\,E^2-M^2)}{2\,\Lambda\,h^2}-M\,\omega-\frac{M\,\omega\,r}{\tan r}-M^2\,\omega^2\,r^2-\frac{k^2}{2\,\Lambda}-\frac{\iota^2}{\sin^2 r}\Bigg]\,\psi=0.\label{e3}
\end{equation}

We solve the above equation by taking an approximation $\sin r \approx r$, $\tan r \approx r$ as done in the previous section. Therefore, the radial wave equation (\ref{e3}) reduces to the following form:
\begin{equation}
    \psi''+\frac{1}{r}\,\psi'+\Bigg[\lambda-M^2\,\omega^2\,r^2-\frac{\iota^2}{r^2}\Bigg]\,\psi=0,\label{e4}
\end{equation}
where we have defined
\begin{equation}
\lambda=\frac{(f^2\,E^2-M^2)}{2\,\Lambda\,h^2}-\frac{k^2}{2\,\Lambda}-2\,M\,\omega.\label{e44}    
\end{equation}

Transforming to a new variable via $s=M\,\omega\,r^2$ into the above equation (\ref{e4}) results the following differential equation form :
\begin{equation}
\psi''(s)+\frac{(c_1-\,c_2\,s)}{s\,(1-c_3\,s)}\,\psi'(s)+\frac{\Big(-\xi_1\,s^2+\xi_2\,s-\xi_3\Big)}{s^2\,(1-c_3\,s)^2}\,\psi (s)=0,\label{e5}
\end{equation}
where $c_1=1$, $c_2=0=c_3$ and 
\begin{equation}
\xi_1=\frac{1}{4},\quad \xi_2=\frac{\lambda}{4\,M\,\omega},\quad \xi_3=\frac{\iota^2}{4}.\label{e6}
\end{equation}

Equation (\ref{e5}) is a homogeneous second-order differential equation which can be solved using the parametric Nikiforov-Uvarov (NU) method \cite{AFN}. Numerous authors have been employed this NU method in solving the quantum mechanical problems in quantum systems (see, for example, \cite{hh4,hh5,hh6,hh7,hh8,hh9}). Therefore, following the approach done in Refs. \cite{hh8, hh9} in the current work, we obtain the approximate relativistic energy eigenvalue relation given by
\begin{equation}
    f^2\,E^2_{n\,m}=M^2+h^2\,\Bigg[k^2+8\,M\,\omega\,\Lambda\,\Bigg(n+1+\frac{|m|}{2\,\alpha}\Bigg)\Bigg]\quad (n=0,1,2,....).\label{e7}
\end{equation}

The corresponding radial wave function will be
\begin{equation}
    \psi_{n\,m} (s)=\mathcal{N}_{n, m}\,s^{\frac{|m|}{2\,\alpha}}\,e^{-s/2}\,L^{\left(\frac{|m|}{\alpha}\right)}_{n} (s),\label{e8}
\end{equation}
where $\mathcal{N}_{n\,m}$ is the normalization constant and $L^{\left(\frac{|m|}{\alpha}\right)}_{n}(x)$ is the generalized/associated  Laguerre polynomial which is orthogonal over $(0, \infty]$ w. r. t. the measure with weighting function $s^{\frac{|m|}{\alpha}}\,e^{-s}$. 

In terms of $r$, we can rewrite the wave function (\ref{e8}) as follows:
\begin{equation}
    \psi_{n\,m} (r)=\mathcal{N}_{n\, m}\,\left(M\,\omega\right)^{\frac{|m|}{2\,\alpha}}\,r^{\frac{|m|}{\alpha}}\,e^{-\frac{1}{2}\,M\,\omega\,r^2}\,L^{\left(\frac{|m|}{\alpha}\right)}_{n} \left(M\,\omega\,r^2\right).\label{special3}
\end{equation}

\subsection{Rainbow function: $f(\chi)=\frac{1}{1-\beta\,\chi},\quad h(\chi)=1$}

{\color{black} In this section, we are mainly interested on the following} pair of the rainbow function given by
\begin{equation}
    f(\chi)=\frac{1}{1-\beta\,\chi},\quad h(\chi)=1.\label{e10}
\end{equation}

Substituting this pair of rainbow function (\ref{e10}) into the relation (\ref{e7}) results 
\begin{equation}
    \frac{E^2}{\Big(1-\beta\,\frac{E}{E_p}\Big)^2}=M^2+k^2+8\,M\,\omega\,\Lambda\,\Big(n+1+\frac{|m|}{2\,\alpha}\Big).\label{e77}
\end{equation}
After simplification of the above equation, we obtain the approximate energy eigenvalue of quantum oscillator fields associated with the mode $\{n, m\}$ given by
\begin{equation}
    E_{n\, m}=\Bigg[\frac{\beta}{E_p} \pm \Bigg\{M^2+k^2+8\,M\,\omega\,\Lambda\,\Big(n+1+\frac{|m|}{2\,\alpha}\Big)\Bigg\}^{-1/2} \Bigg]^{-1}.\label{e11}
\end{equation}

\begin{figure}[ht!]
\subfloat[$\alpha=0.9,n=1=m=\omega$]{\centering{}\includegraphics[scale=0.55]{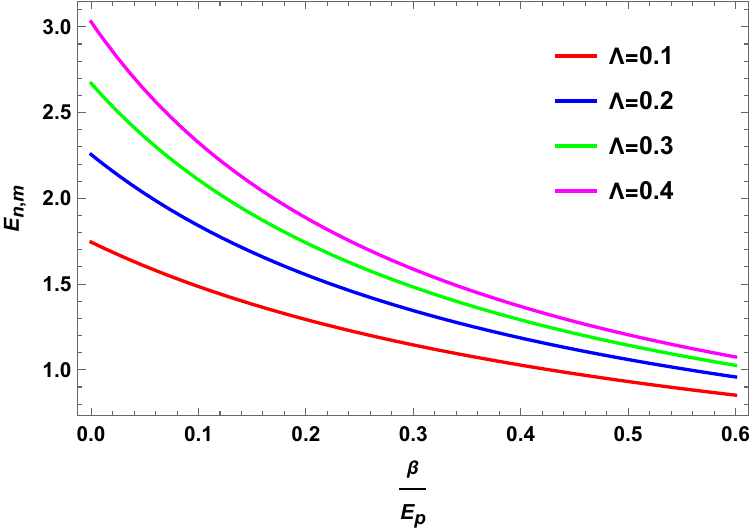}}\quad\quad
\subfloat[$\Lambda=0.1,n=1=m=\omega$]{\centering{}\includegraphics[scale=0.55]{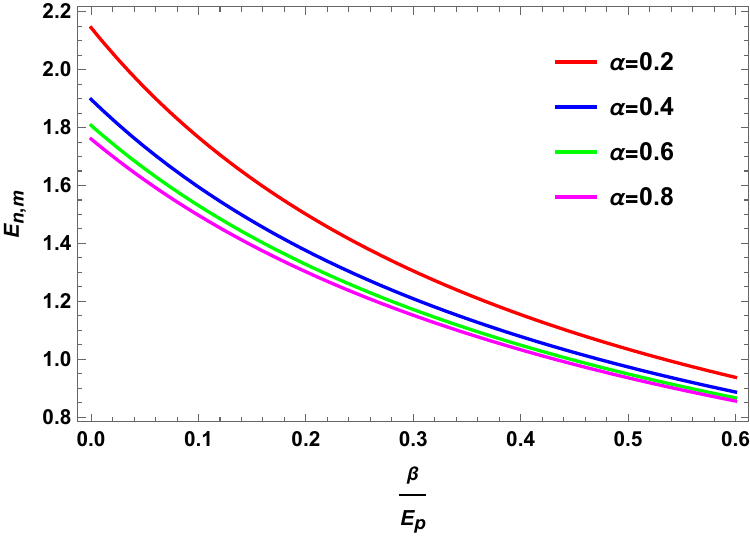}}\\
\subfloat[$\alpha=0.9,\Lambda=0.1,m=1=\omega$]{\centering{}\includegraphics[scale=0.55]{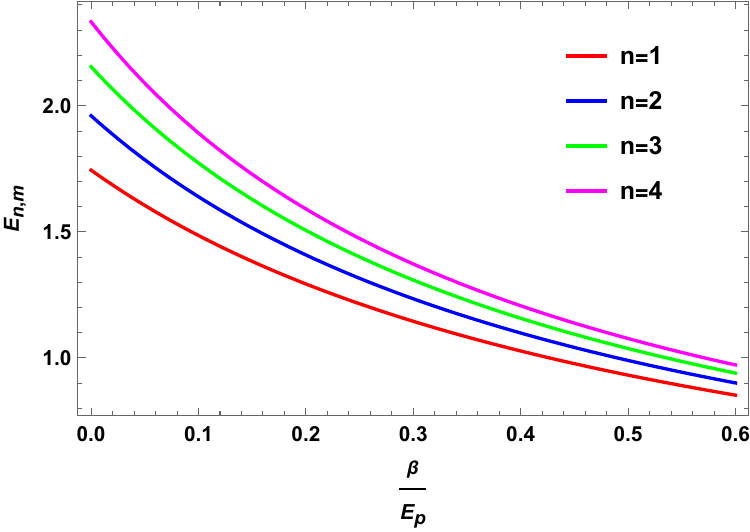}}\quad\quad
\subfloat[$\alpha=0.9,\Lambda=0.1,n=1=m$]{\centering{}\includegraphics[scale=0.55]{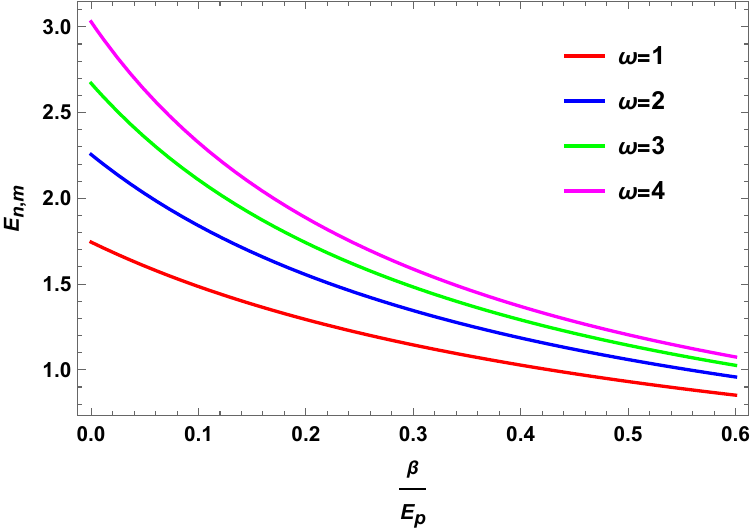}}\\
\subfloat[$\alpha=0.9,m=1=\omega$]{\centering{}\includegraphics[scale=0.55]{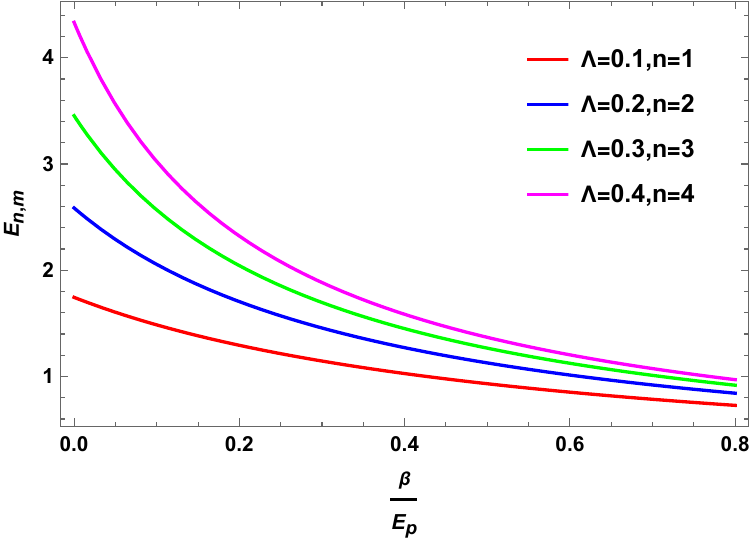}}\quad\quad\quad
\subfloat[$\Lambda=0.1,n=1=m$]{\centering{}\includegraphics[scale=0.55]{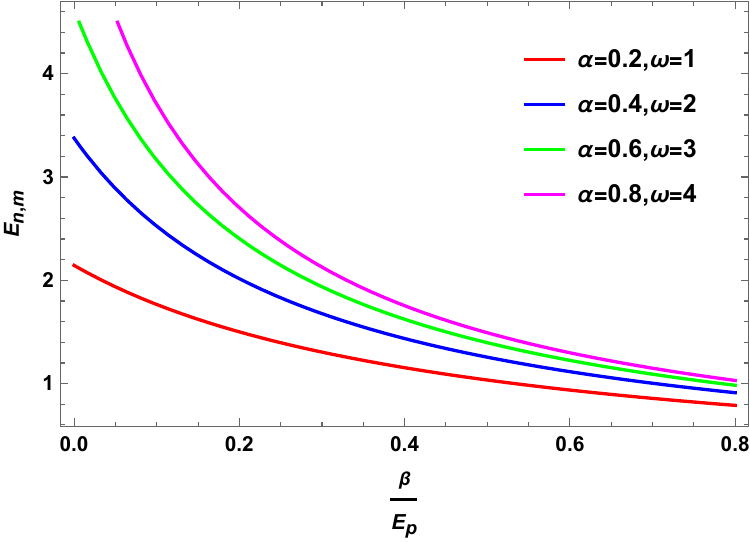}}
\centering{}\caption{The energy spectrum $E_{n\,m}^{+}$ for Eq. (\ref{e11}) taking plus sign within the bracket, where parameters are set as $M=1$ and $k=0$.}
\label{figure:4}
\end{figure}

\begin{figure}[ht!]
\subfloat[$\alpha=0.9,n=1=m=\omega$]{\centering{}\includegraphics[scale=0.55]{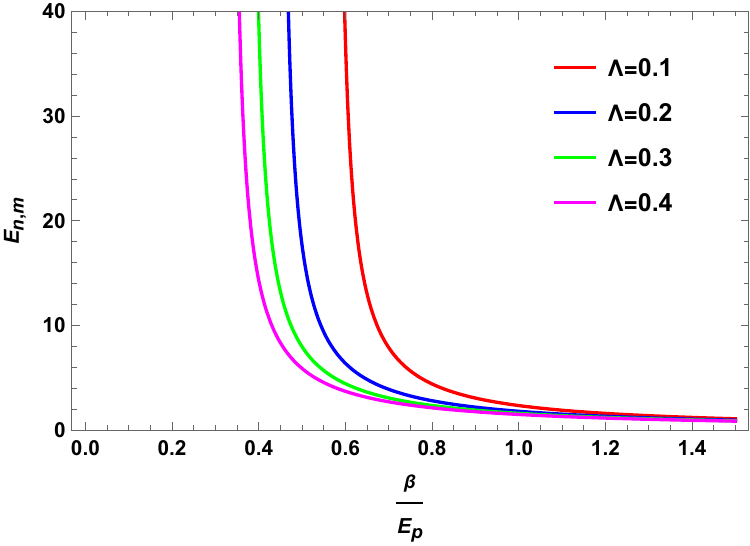}}\quad\quad
\subfloat[$\Lambda=0.1,n=1=m=\omega$]{\centering{}\includegraphics[scale=0.55]{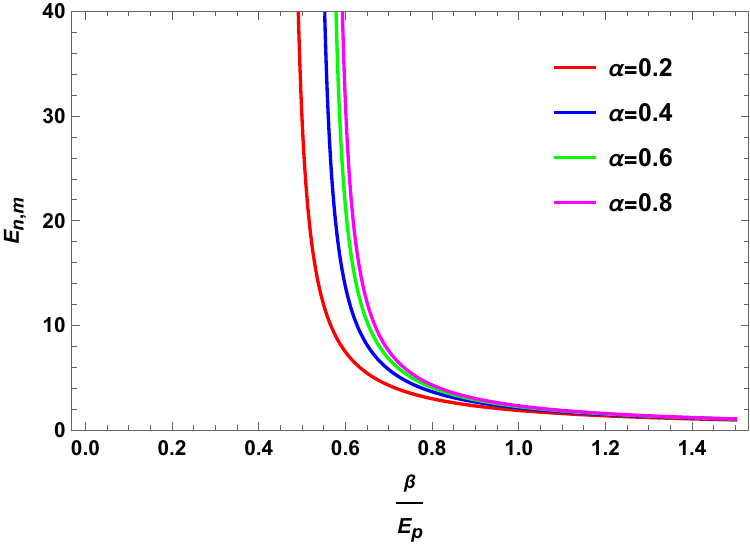}}\\
\subfloat[$\alpha=0.9,\Lambda=0.1,m=1=\omega$]{\centering{}\includegraphics[scale=0.55]{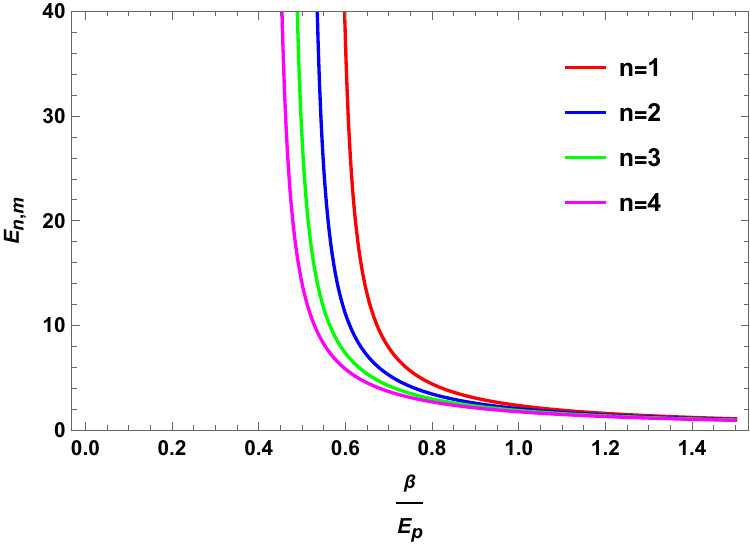}}\quad\quad
\subfloat[$\alpha=0.9,\Lambda=0.1,n=1=m$]{\centering{}\includegraphics[scale=0.55]{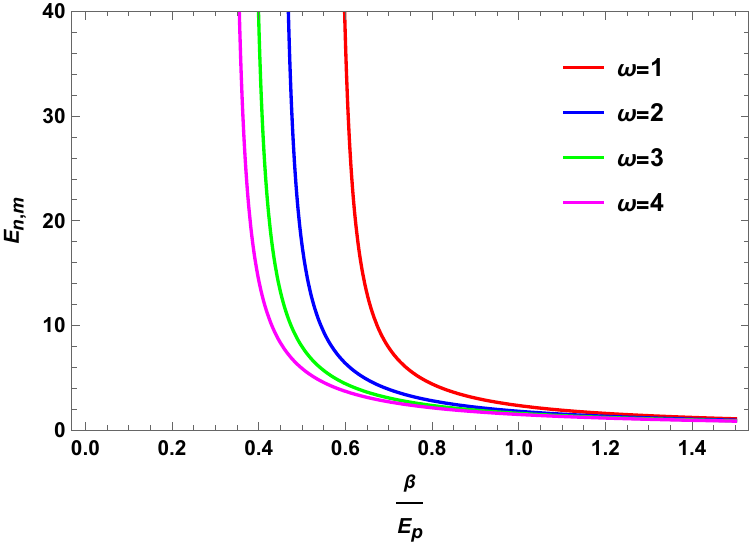}}\\
\subfloat[$\alpha=0.9,m=1=\omega$]{\centering{}\includegraphics[scale=0.55]{Equation-35-Figure-1-e.pdf}}\quad\quad
\subfloat[$\Lambda=0.1,n=1=m$]{\centering{}\includegraphics[scale=0.55]{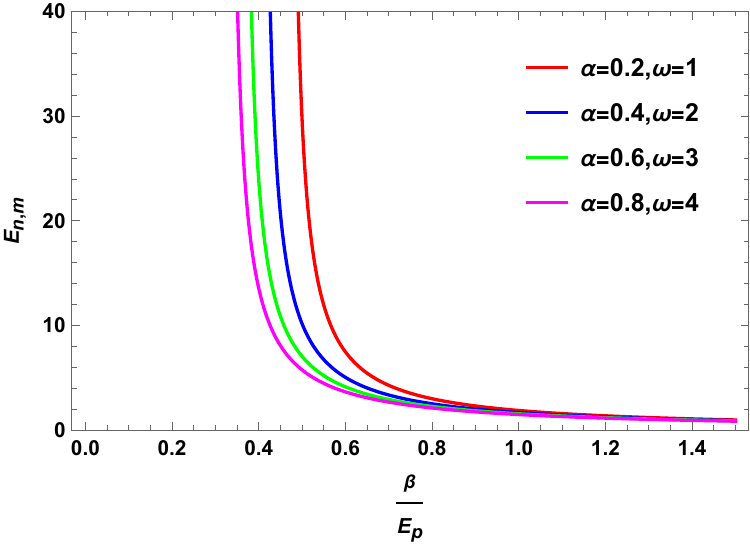}}
\centering{}\caption{The energy spectrum $E_{n\,m}$ of Eq. (\ref{e11}) taking minus sign within the bracket, where parameters are set as $M=1$, $k=0$.}
\label{figure:5}
\end{figure}

Equation (\ref{e11}) is the approximate relativistic energy profile of oscillator fields in BM-universe with a nonzero cosmological under the influence of rainbow gravity. We see that the approximate relativistic energy spectrum and the radial wave function is influenced by the topology of the geometry characterized by the parameter $\alpha$, the cosmological constant $\Lambda$, and the rainbow parameter $\beta$. Moreover, these are changes in the quantum numbers $\{n,m\}$. 

{\color{black} In the limit $\beta =0$, i. e. there is no rainbow-gravity effect, the energy eigenvalue (\ref{e11}) reduces to the following form:
\begin{equation}
    E_{n\, m}=\sqrt{M^2+k^2+8\,M\,\omega\,\Lambda\,\Bigg(n+1+\frac{|m|}{2\,\alpha}\Bigg)}.\label{e12}
\end{equation}

Equation (\ref{e12}) is the approximate relativistic energy profile of quantum oscillator fields in BM-universe with a cosmological without rainbow gravity effects.

Thus, by comparing Eqs. (\ref{e11}) and (\ref{e12}), it becomes evident that the presence of rainbow gravity alters the relativistic energy levels of quantum oscillator fields when compared to the scenario without this effect.

}

\begin{figure}
\subfloat[$n=0$, $\omega=1$]{\centering{}\includegraphics[scale=0.55]{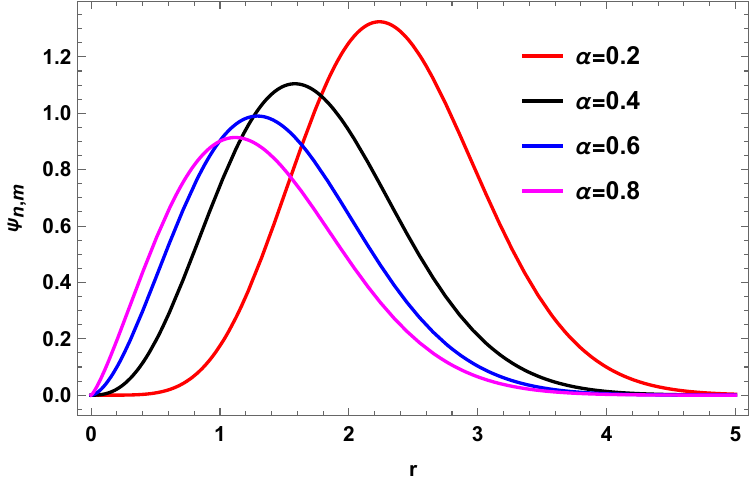}}\quad\quad\quad
\subfloat[$n=0$, $\alpha=0.5$]{\centering{}\includegraphics[scale=0.55]{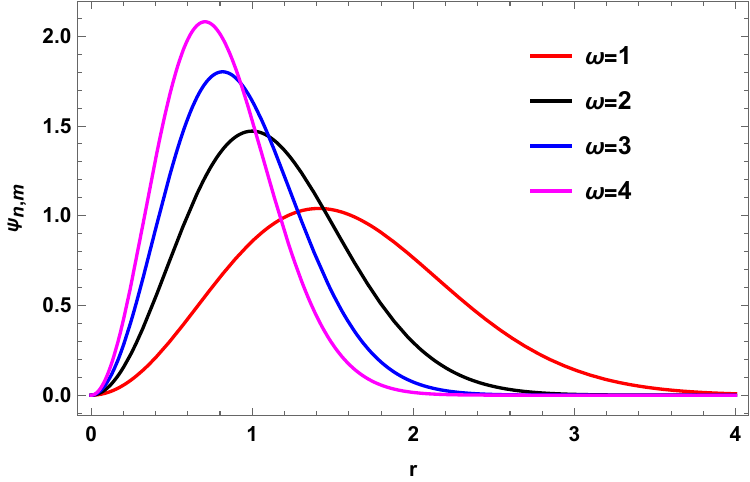}}\\
\subfloat[$\alpha=0.5$, $\omega=1$]{\centering{}\includegraphics[scale=0.55]{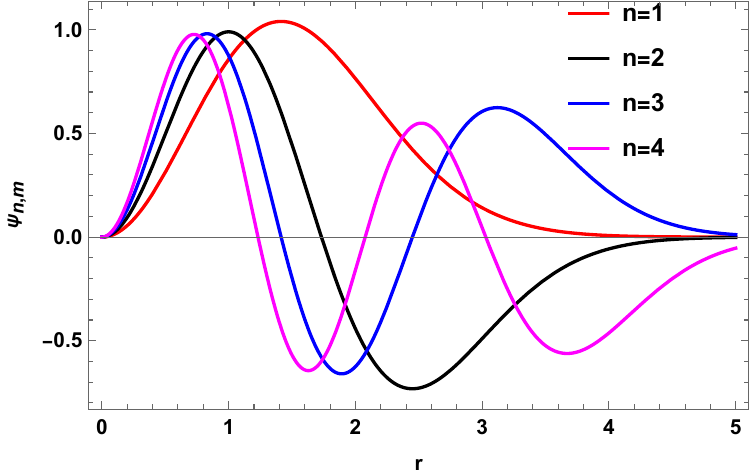}}\quad\quad\quad
\subfloat[$n=0$]{\centering{}\includegraphics[scale=0.55]{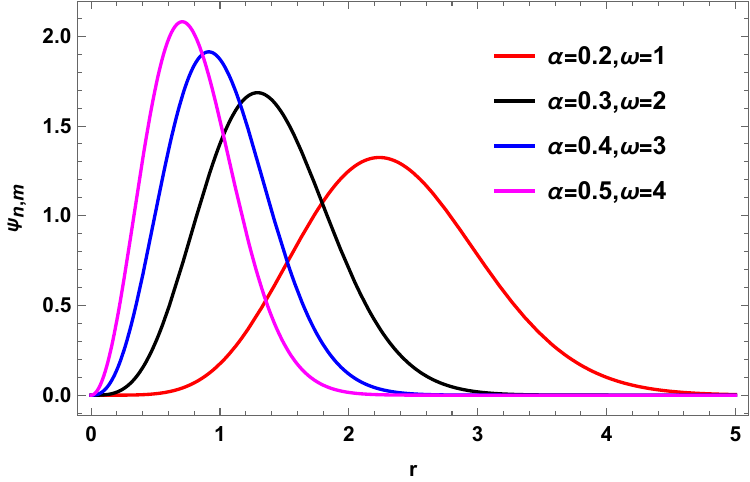}}
\centering{}\caption{The radial wave function $\psi_{n\,m} (r)$ of Eq. (\ref{special3}), where parameters are set as $M=1=m$ and $\Lambda=0.5$.}
\label{figure:6}
\hfill\\
\subfloat[$n=1$, $\omega=1$]{\centering{}\includegraphics[scale=0.55]{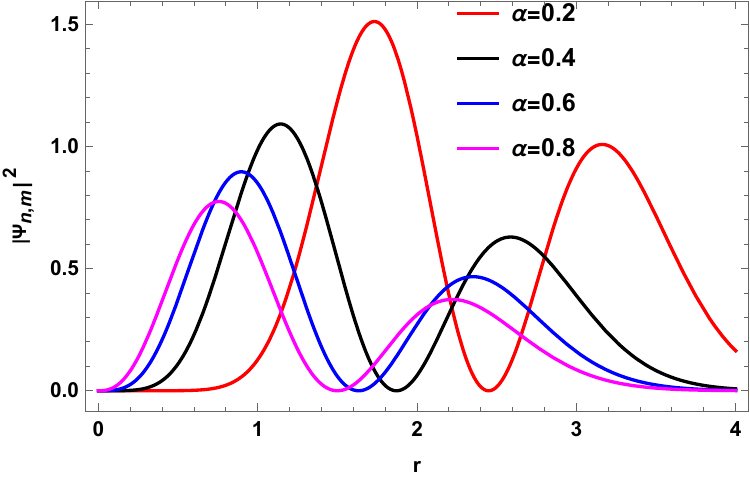}}\quad\quad\quad
\subfloat[$n=1$, $\alpha=0.5$]{\centering{}\includegraphics[scale=0.55]{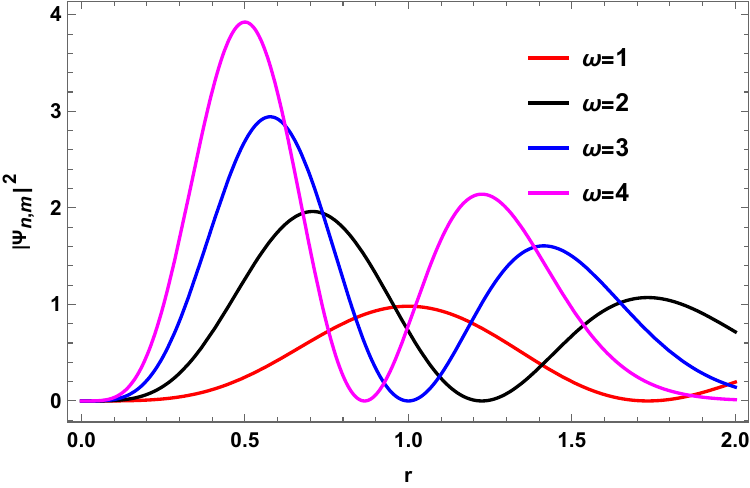}}\\
\subfloat[$\alpha=0.5$, $\omega=1$]{\centering{}\includegraphics[scale=0.55]{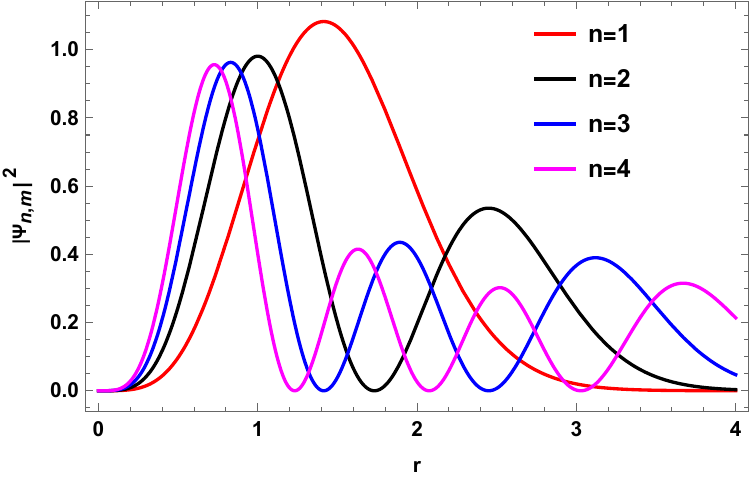}}\quad\quad\quad
\subfloat[$n=1$]{\centering{}\includegraphics[scale=0.55]{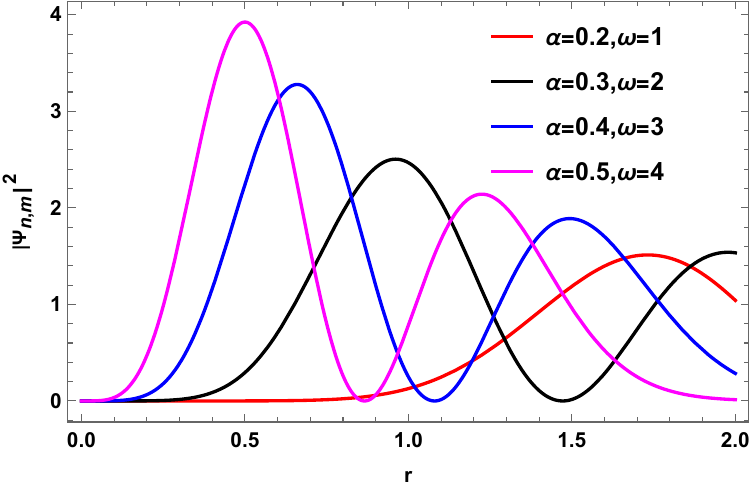}}
\centering{}\caption{The probability density function $|\Psi_{n\,m}(r)|^2$, where parameters are set as $M=1=m$ and $\Lambda=0.5$.}
\label{figure:7}
\end{figure}

We have presented Figures \ref{figure:4}--\ref{figure:5}, displaying the energy spectrum $E_{n,m}$ of expression (\ref{e11})) plotted against $\beta/E_p$. The figure encompasses various values of the cosmological constant $\Lambda$ (Fig: 4(a) \& 5(a)), the topology parameter $\alpha$ (Fig: 4(b) \& 5(b)), the radial quantum number $n$ (Fig: 4(c) \& 5(c)), the oscillator frequency $\omega$ (Fig: 4(d) \& 5(d)), both the cosmological constant $\Lambda$ and quantum number $n$ (Fig: 4(e) \& 5(e)), and both the topological parameter $\alpha$ and the oscillator frequency $\omega$ (Fig: 4(f) \& 5(f)). From these Figures \ref{figure:4}--\ref{figure:5}, we see that for $\beta/E_p = 0$ implies $\beta=0$, the energy eigenvalue $E$ possesses some finite values other than zero. {\color{black} It is noted that by setting specific values for the various parameters, we have observed that at $\beta=0$, the energy eigenvalue $E_{n\,m}$ of Eq. (\ref{e11}) becomes finite and very large. In the graphical representations (Figure \ref{figure:5}), we have considered this value within a particular range, for instance, $0 <E_{n\,m} <40$. This is why some figures in \ref{figure:5} exhibit divergence behavior but actually it is not. However, extending the range beyond this considered limit would result in graphical representations that are not as clear.}

We have generated Figure \ref{figure:6}, which illustrates the nature of the normalized radial wave function $\psi_{n\,m}(r)$. The figure includes variations for different values of the topology parameter $\alpha$ (Fig. 6(a)), the oscillator frequency $\omega$ (Fig. 6(b)), the radial quantum number $n$ (Fig. 6(c)), and a combination of $\alpha$ and $\omega$ (Fig. 6(d)).

Figure \ref{figure:7} depicts the probability density function $P(r)=\Psi^{*}_{n,m}\,\Psi_{n\,m}=|\Psi_{n\,m}|^2$, highlighting the effects of varying the topology parameter $\alpha$ (Fig. 7(a)), the oscillator frequency $\omega$ (Fig. 7(b)), the radial quantum number $n$ (Fig. 7(c)), and a combination of $\alpha$ and $\omega$ (Fig. 7(d)). In both figures, we maintain a quantum number $m=1$ and set the cosmological constant $\Lambda=0.5$. These figures collectively illustrate how changes in the topology parameter, oscillator frequency, and radial quantum number influence the radial wave function and the corresponding probability density function.

\section{Conclusions}\label{sec:4}

In this analysis, we primarily focused on the relativistic quantum dynamics of charge-free scalar particles, governed by the Klein-Gordon equation, within a curved space-time background (\ref{a1}). This curved geometry is derived from an Einstein-Maxwell equation of Bonnor-Melvin solution \cite{BB1,BB2}, which incorporates a positive cosmological constant. Subsequently, we studied the quantum dynamics of an oscillator field described by the Klein-Gordon oscillator equation within the same curved space-time framework. It is important to note that this entire investigation of quantum mechanical problems is conducted within the rainbow-gravity environment, which deformed the space-time geometry. 

In Section \ref{sec:2}, we derived the radial equation of the wave equation under the influence of RG and obtained an approximate relativistic energy eigenvalue expression using special functions. To achieve this, we considered two different types of rainbow functions that have applications in loop quantum gravity and related fields. For these pairs of rainbow functions, the approximate relativistic energy profiles of charge-free scalar particles were derived and are presented in Eqs. (\ref{c5}) and (\ref{c6}). We demonstrated that these energy spectra are influenced by various factors, such as the cosmological constant ($\Lambda$), the topology parameter ($\alpha$), and the rainbow parameter ($\beta$) and changes in the quantum numbers ($n$, $m$). To illustrate their influence on the eigenvalues, we generated several graphs (Figures 1--3) depicting the nature of the energy spectra for different values of these parameters.   

In Section \ref{sec:3}, we shifted our focus to the quantum dynamics of an oscillator field within the same magnetic universe, considering the influence of RG. We derived the radial equation of the Klein-Gordon oscillator equation and presented the approximate relativistic energy eigenvalues using special functions. To achieve this, we employed a specific pair of rainbow functions given in Eq. (\ref{e10}) and obtained the approximate energy spectrum presented in Eq. (\ref{e11}). We observed that the relativistic approximate energy spectrum of the oscillator field is subject to modifications influenced by several factors: the topology parameter ($\alpha$), the cosmological constant ($\Lambda$), and the rainbow parameter ($\beta$). Additionally, the oscillator frequency ($\omega$) contributes to alterations in this relativistic energy profile, along with changes in the quantum numbers ($n$, $m$). To illustrate the energy spectrum, we generated Figures 4 and 5, which show the influence of the cosmological constant, the topology parameter, and the quantum numbers. These figures reveal some interesting results and patterns in the behavior of the energy spectrum under varying conditions.

{\color{black} 
It is worth mentioning that the quantum mechanical problems, specifically the Klein-Gordon equation in the context of the original Bonnor-Melvin solution \cite{BB1,BB2}, has not been studied in the literature to the best of our knowledge. Therefore, it is not possible to compare our results with those results. However, in our future studies, we plan to address quantum mechanical problems within the context of Bonnor-Melvin solution \cite{BB1,BB2}.
}

{\color{black} 
In summary, our research explored quantum field theory in curved space-time within the rainbow gravity framework. We found intriguing results on the behavior of charge-free scalar particles and quantum oscillator fields in this setting, particularly under the influence of rainbow gravity. These findings deepen our understanding of the interplay between quantum fields, curved space-time, and rainbow gravity effects. In our future work, we plan to extend our analysis to study quasi-normal modes (QNMs), following the methodology outlined in Refs. \cite{t1, t2}. We will then analyze the resulting outcomes.}

\section*{Acknowledgements}

We sincerely acknowledges the anonymous referees for their valuable comments and helpful suggestions. F. A acknowledges the Inter University Centre for Astronomy and Astrophysics (IUCAA), Pune, India for granting visiting associateship.

\section*{Conflict of Interest}

There is no conflict of interests.

\section*{Funding Statement}

There is no funding agency associated with this manuscript.

\section*{Data Availability Statement}

No data were generated or analysed during this study.


\begin{thebibliography}{}

\bibitem{k1} A. Einstein, Ann. Phys. (Berlin) {\bf 354}, 769 (1916).

\bibitem{k2} B. P. Abbott {\it et al.}, Phys. Rev. Lett. {\bf 116 }, 061102 (2016).

\bibitem{k3} K. Akiyama {\it et al.}, Astrophys. J. Lett. {\bf 875}, L1 (2019).

{\color{black}{
\bibitem{t1} Y. Zhao, X. Ren, A. Ilyas,  E. N. Saridakis and Y. F. Cai, JCAP {\bf 2022}(10), 087 (2022).

\bibitem{t2} Y. Zhao, Y. Cai, S. Das, G. Lambiase, E. N. Saridakis and E. C. Vagenas, Nucl. Phys. {\bf B 1004}, 116545 (2024).
}}

\bibitem{k4} R. P. Feynman and A. R. Hibbs, {\it Quantum Mechanics and Path Integrals}, Dover Publications Inc. (1965).

\bibitem{k5} G. Amelino-Camelia, Phys. Lett. \textbf{B 510}, 255 (2001).

\bibitem{k6} G. Amelino-Camelia, Int. J. Mod. Phys. \textbf{D 11}, 35 (2002). 

\bibitem{k7} G. Amelino-Camelia, J. Kowalski-Glikman, G. Mandanici, and A. Procaccini, Int. J. Mod. Phys. \textbf{A 20}, 6007 (2005).

\bibitem{k8} J. Magueijo, and L. Smolin, Phys. Rev. Lett. \textbf{88}, 190403 (2002). 

\bibitem{k9} J. Magueijo, and L. Smolin, Phys. Rev. \textbf{D 67}, 044017 (2003).

\bibitem{kk10} J. Magueijo, and L. Smolin,  Class. Quantum Grav. {\bf 21}, 1725 (2004).

\bibitem{k10} L. Smolin, Nucl. Phys. \textbf{B 742}, 142 (2006).

\bibitem{k11} S. Ghosh, Phys. Rev. \textbf{D 74}, 084019 (2006).

\bibitem{k12} Y. Ling, and Q. Wu, Phys. Lett. \textbf{B 687}, 103 (2010).

\bibitem{k13} A. Ashour, M. Faizal, A. F. Ali, and F. Hammad, Eur. Phys. J. \textbf{C 76}, 264 (2016).

\bibitem{k14} G. Amelino-Camelia, J. R. Ellis, N. E. Mavromatos, D. V. Nanopoulos, and S. Sarkar, Nature \textbf{393}, 763 (1998).

\bibitem{k15} G. Amelino-Camelia, J. R. Ellis, N. E. Mavromatos, and D. V. Nanopoulos, Int. J. Mod. Phys. \textbf{A 12}, 607 (1997).

\bibitem{k16} V. A. Kostelecký, and S. Samuel, Phys. Rev. \textbf{D 39}, 683 (1989).

\bibitem{k17} R. Gambini, and J. Pullin, Phys. Rev. \textbf{D 59}, 124021 (1999).

\bibitem{k18} G. T. Hooft, Class. Quantum Grav. \textbf{13}, 1023 (1996).

\bibitem{k19} S. M. Carroll, J. A. Harvey, V. A. Kostelecký, C. D. Lane, and T. Okamoto, Phys. Rev. Lett. \textbf{87}, 141601 (2001).

\bibitem{pp2} J. Magueijo, and L. Smolin, Phys. Rev. {\bf D 71}, 026010 (2005).

\bibitem{k20} B. Majumder, Int. J. Mod. Phys. \textbf{D 22}, 1350079 (2013).

\bibitem{k21} S. H. Hendi, M. Momennia, B. Eslam-Panah, and S. Panahiyan, Phys. Dark Univ. \textbf{16}, 26 (2017).

\bibitem{k22} S. H. Hendi, M. Faizal, B. Eslam-Panah, and S. Panahiyan, Eur. Phys. J. \textbf{C 76}, 296 (2016).

\bibitem{k23} C. Leiva, J. Saavedra, and J. Villanueva, Mod. Phys. Lett. \textbf{A 24}, 1443 (2009).

\bibitem{k24} H. Li, Y. Ling, and X. Han, Class. Quantum Grav. \textbf{26}, 065004 (2009).

\bibitem{k25} V. B. Bezerra, H .F. Mota, and C. R. Muniz, EPL \textbf{120}, 10005 (2017).

\bibitem{k26} R. Garattini, and G. Mandanici, Eur. Phys. J. \textbf{C 77}, 57 (2017).

\bibitem{k27} R. Garattini, and E. N. Saridakis, Eur. Phys. J. \textbf{C 75}, 343 (2015).

\bibitem{k28} R. Garattini, and F. S. N. Lobo, Eur. Phys. J. \textbf{C 74}, 2884 (2014).

\bibitem{k29} R. Garattini, and B. Majumder, Nucl. Phys. \textbf{B 883}, 598 (2014).

\bibitem{k30} R. Garattini, JCAP 06 ({\bf 2013}) 017.

\bibitem{k31} R. Garattini, Phys. Lett. \textbf{B 685}, 329 (2010).

\bibitem{k32} R. Garattini, and G. Mandanici, Phys. Rev. \textbf{D 83}, 084021 (2011).

\bibitem{k33} R. Garattini, and G. Mandanici, Phys. Rev. \textbf{D 85}, 023507 (2012).

\bibitem{k34} R. Garattini, and B. Majumder, Nucl. Phys. \textbf{B 884}, 125 (2014).

\bibitem{k35} R. Garattini, and M. Sakellariadou, Phys. Rev. \textbf{D 90}, 043521 (2014).

\bibitem{k36} R. Garattini, and F. S. N. Lobo, Phys. Rev. \textbf{D 85}, 024043 (2012).

\bibitem{k37} A. F. Ali, M. Faizal, and B. Majumder, EPL \textbf{109}, 20001 (2015).

\bibitem{k38} A. F. Ali, M. Faizal, B. Majumder, and R. Mistry, Int. J. Geom. Meths. Mod. Phys. \textbf{12}, 1550085 (2015).

\bibitem{BEP1} B. Eslam Panah, Phys. Lett. {\bf B 787}, 45 (2018).

\bibitem{BEP2} B. Eslam Panah, G. H. Bordbar, S. H. Hendi, R. Ruffini, Z. Rezaei, and R. Moradi, Astrophys. J. {\bf 848}, 24 (2017).

\bibitem{BEP3} S. H. Hendi, G. H. Bordbar, B. Eslam Panah, and S. Panahiyan, JCAP 09 ({\bf 2016}) 013.

\bibitem{BEP4} A. B. Tudeshki, G. H. Bordbar, and B. Eslam Panah, Phys. Lett. {\bf B 835}, 137523 (2022).

\bibitem{BEP5} A. B. Tudeshki, G. H. Bordbar, and B. Eslam Panah, Phys. Dark Universe. {\bf 42}, 101354 (2023).

\bibitem{BEP6} H. Barzegar, M. Bigdeli, G. H. Bordbar, and B. Eslam Panah, Eur. Phys. J C {\bf 83}, 151 (2023).

\bibitem{BEP7} A. B. Tudeshki, G. H. Bordbar, and B. Eslam Panah, Phys. Lett. {\bf B 848}, 138333 (2024).

\bibitem{kk13} C. Thompson, and R. C. Duncan, MNRAS {\bf 275}, 255 (1995).

\bibitem{kk14} C. Kouveliotou {\it et al.}, Nature {\bf 393}, 235 (1998).

\bibitem{kk15} U. Gürsoy, D. Kharzeev, and K. Rajagopal, Phys. Rev. {\bf C 89}, 054905 (2014).

\bibitem{kk16} A. Bzdak, and V. Skokov, Phys. Lett. {\bf B 710}, 171 (2012).

\bibitem{kk17} V. Voronyuk, V. D. Toneev, W. Cassing, E. L. Bratkovskaya, V. P. Konchakovski, and S. A. Voloshin, Phys. Rev. {\bf C 83}, 054911 (2011).

\bibitem{kk18} T. Gutsunaev, and V. Manko, Phys. Lett. {\bf A 123}, 215 (1987).

\bibitem{kk19} T. Gutsunaev, and V. Manko, Phys. Lett. {\bf A 132}, 85 (1988).

\bibitem{kk20} W. B. Bonnor, Proc. Phys. Soc. {\bf A 67}, 225 (1954).

\bibitem{kk21} M. Melvin, Phys. Lett. {\bf 8}, 65 (1964).

\bibitem{MZ2} M. Astorino, JHEP 06 ({\bf 2012}) 086.

\bibitem{MZ3} J. Vesely, and M. \v{Z}ofka, Phys. Rev. {\bf D 100}, 044059 (2019).

\bibitem{kk22} M. Žofka, Phys. Rev. {\bf D 99}, 044058 (2019).

\bibitem{kk23} L. Parker, Phys. Rev. Lett. {\bf 44}, 1559 (1980). 

\bibitem{kk24} N. D. Birrell, and P. C. W. Davies, {\it Quantum fields in curved space},  Cambridge University Press, Cambridge (2012).

\bibitem{kk25} E. Elizalde, Phys. Rev. {\bf D 36}, 1269 (1987).

\bibitem{kk26} S. Chandrasekhar, Proc. R. Soc. London. A: Math. Phys. Sci. {\bf 349}, 571 (1976).

\bibitem{kk27} L. C. N. Santos, and C. C. Barros Jr., Eur. Phys. J. C {\bf 77}, 186 (2017).

\bibitem{kk28} L. C. N. Santos, and C. C. Barros Jr., Eur. Phys. J. C {\bf 78}, 13 (2018).

\bibitem{FA} F. Ahmed, and A. Guvendi, Chin. J. Phys. {\bf 87}, 174 (2024).

\bibitem{kk32} L. C. N. Santos, C. E. Mota, and C. C. Barros Jr., Adv. High Energy Phys. {\bf 2019}, 2729352 (2019).

\bibitem{kk35} A. Bouzenada, and A. Boumali, Ann. Phys. (NY) \textbf{452}, 169302 (2023).

\bibitem{kk35-1} A. Bouzenada, A. Boumali, R. L. L. Vitoria, F. Ahmed, and M. Al-Raeei, Nucl. Phys. \textbf{B 994}, 116288 (2023).

\bibitem{kk35-2} A. Bouzenada, A. Boumali, and F. Serdouk, Theor. Math. Phys. \textbf{216}, 1055 (2023).

\bibitem{kk35-3} A. Bouzenada, A. Boumali, and E. O. Silva, Ann. Phys. (NY) \textbf{458}, 169479 (2023).

\bibitem{AG} F. Ahmed, and A. Guvendi,  Nucl. Phys. {\bf B 1000}, 116470 (2024).

\bibitem{kk36}V. B. Bezerra, M. S. Cunha, L. F. F. Freitas, C. R. Muniz, and M. O. Tahim, Mod. Phys. Lett. {\bf A 32}, 1750005 (2016).

\bibitem{kk37} L. C. N. Santos, and C. C. Barros Jr., Int. J. Mod. Phys. {\bf A 33}, 1850122 (2018).

\bibitem{kk38} E. O. Pinho, and C. C. Barros Jr., Eur. Phys. J. {\bf C 83}, 745 (2023).

\bibitem{kk39} P. Sedaghatnia, H. Hassanabadi, and F. Ahmed, Eur. Phys. J. {\bf C 79}, 541 (2019).

\bibitem{kk40} A. Guvendi, S. Zare, and H. Hassanabadi, Phys. Dark Univ. {\bf 38}, 101133 (2022). 

\bibitem{kk42} L. C. N. Santos, and C. C. Barros Jr., Eur. Phys. J.  {\bf C 76}, 560 (2016).

\bibitem{AB2} F. Ahmed, and A. Bouzenada, Commun. Theor. Phys. {\bf 76}, 045401 (2024).

\bibitem{AB3} F. Ahmed, and A. Bouzenada, Nucl. Phys. {\bf B 1000}, 116490 (2024).

\bibitem{AB4} F. Ahmed, and A. Bouzenada, Phys. Scr. {\bf 99}, 065033 (2024).

\bibitem{ZWF} Z. W. Feng, and S. Z. Yang, Phys. Lett. {\bf B 772}, 737 (2017).

\bibitem{SHH} S. H. Hendi, M. Faizal, B. Eslam Panah and S. Panahiyan, Eur. Phys. J C {\bf 76}, 296 (2016).

\bibitem{WG} W. Greiner, {\it Relativistic Quantum Mechanics. Wave Equations},  Springer-Verlag, Berlin, Germany (2000).

\bibitem{MA} M. Abramowitz, and I. A. Stegun, {\tt Handbook of Mathematical Functions with Formulas, Graphs, and Mathematical Tables}, New York: Dover (1972).

\bibitem{GBA} G. B. Arfken, H. J. Weber, and F. E. Harris, {\it Mathematical Methods for Physicists}, Elsevier (2012).

\bibitem{AFN} A. F. Nikiforov, and V. B. Uvarov, {\it Special Functions of Mathematical Physics}, Birkhauser (1988).

\bibitem{hh4} H. Hassanabadi, E. Maghsoodi, and S. Zarrinkamar, H. Rahimov, Chin. Phys. {\bf B 21}, 120302 (2012).   

\bibitem{hh5} H. Hassanabadi, E. Maghsoodi, and S. Zarrinkamar, H. Rahimov, J. Math. Phys. {\bf 53}, 022104 (2012).

\bibitem{hh6} E. Maghsoodi, H. Hassanabadi, and S. Zarrinkamar, Chin. Phys. {\bf B 22}, 030302 (2013).

\bibitem{hh7} H. Hassanabadi, E. Maghsoodi, A. N. Ikot, and S. Zarrinkamar, Appl. Math. Comput. {\bf 219}, 9388 (2013).

\bibitem{hh8} M. de Montigny, H. Hassanabadi, J. Pinfold, and  S. Zare, Eur. Phys. J. Plus {\bf 136}, 788 (2021).

\bibitem{hh9} M. de Montigny, J. Pinfold, S. Zare, and  H. Hassanabadi, Eur. Phys. J. Plus {\bf 137}, 54 (2022).

\bibitem{BB1} W. B. Bonnor, Proc. Phys. Soc. London Sect. {\bf A 67}, 225 (1954).

\bibitem{BB2} M. A. Melvin, Phys. Lett. {\bf 8}, 65 (1964).



\end{thebibliography}
\end{document}